\documentclass[MSNbibl,nameyear,seceqn,dvips]{arxstspdf}
\usepackage{flushend}
\usepackage{stfloats}

\volume{29}
\issue{3}
\pubyear{2014}
\firstpage{359}
\lastpage{362}
\doi{10.1214/14-STS494} 
\referstodoi{10.1214/14-STS480}
\docsubty{FLA}



\begin{document}
\begin{frontmatter}\vspace*{12pt}

\title{Instrumental Variables Before and LATEr}
\runtitle{Comment}

\begin{aug}
\author[a]{\fnms{Toru} \snm{Kitagawa}\ead[label=e1]{t.kitagawa@ucl.ac.uk}}
\runauthor{T. Kitagawa}

\affiliation{University College London}

\address[a]{Toru Kitagawa is Lecturer, Department of Economics,
University College London, London, WC1E 6BT, United Kingdom \printead{e1}.}
\end{aug}

%
\begin{abstract}
The modern formulation of the instrumental variable methods initiated
the valuable interactions between economics and statistics literatures
of causal inference and fueled new innovations of the idea. It helped
resolving the long-standing confusion that the statisticians used to
have on the method, and encouraged the economists to rethink how to
make use of instrumental variables in policy analysis.
\end{abstract}

%
\end{frontmatter}

It is an honor to comment on Professor Imbens' paper on instrumental
variables methods. The discussed paper reviews both the origin of the
instrumental variables methods in econometrics and their modern formulation
and interpretation based on the concept of potential outcomes
originating in
statistics. A unique feature of this review article is its comparative
perspective. Imbens convinces us that ``choice versus chance in
treatment assignment'' best summarizes the difference between econometrics
and statistics in their traditions of identifying causal effects.

The seminal papers by Angrist, Imbens and Rubin (\cite{ImbensandAngrist1994}; \cite{AIR}) on the potential outcome-based
formulation of the instrumental variables method are some of the few rare
works that generated equally enormous influence on both econometrics and
statistics communities. In the economics side, the major impacts appear in
the following three aspects. First, the modern way of viewing an
instrumental variable in relation to treatment noncompliance and an
encouragement design widened the scope of applications of the method.
Traditionally, the uses of the instrumental variables method were restricted
to observational studies, and economic theories or researcher's background
knowledge on the problem were playing a unique role in validating the
exogeneity and exclusion restrictions of the employed instrument.
Nowadays, this new encouragement design viewpoint offers another strategy
for finding an instrument in a given application, and with a randomized
initial treatment assignment, researchers can validate easily and credibly
the instrument exogeneity assumption without resorting to an economic
theory. Second, the concept of the local average treatment effect
considerably changed the way we interpret the estimation results. We are
no longer puzzled by obtaining contradicting estimation results across
different instruments, and we treat them as separate and valuable
pieces of
information about heterogeneous causal effects. In addition, acknowledging
nonidentfiability of the population average causal effect has promoted the
discussion of whether the instrumental variable method should be used for
the actual policy decision making and how. Third, the discovery of the
importance of instrument monotonicity assumption led us to think more
carefully about the subjects' causal/behavioral responses to the assigned
instrument.

In what follows, I first illustrate by an example the link
between the
textbook linear instrumental variable model and the potential outcome
framework to complement the discussion that Imbens gave in Section~6. In
the second part, I review the active but unsettled discussions about
usefulness of estimating the local average treatment effect, and provide
briefly my personal opinion on the issue.

\section{Causal Interpretation in the Textbook Model}

The standard econometrics education introduces the instrumental variables
methods in the form of, what Imbens called, the standard textbook set
up,%
%
\begin{equation}\label{textbook}
Y_{i}^{\mathrm{obs}}=\beta_{0}+\beta_{1}X_{i}^{\mathrm{obs}}+
\beta_{2}^{\prime
}V_{i}+\varepsilon_{i},
\end{equation}
where $Y_{i}^{\mathrm{obs}}$ is an outcome observation of unit $i$,
$X_{i}^{\mathrm{obs}}$ is
a treatment variable of which the causal effect on the outcome is of
interest, $V_{i}$ is a vector of observable covariates (often called control
covariates), and $\varepsilon_{i}$ is an unobservable term often called
as an
unobserved heterogeneity of unit $i$. A common way to motivate the use of
instrumental variables is by invalidating the least square method due to
``the correlation between $X_{i}^{\mathrm{obs}}$ and $\varepsilon_{i}$.'' This quick
but somewhat less rigorous way of motivating the instrumental variables
methods often creates confusions. If equation (\ref{textbook}) were
specifying a regression equation or a linear projection, then the projection
residual $\varepsilon_{i}$ is by construction uncorrelated with $X_{i}^{\mathrm{obs}}$,
and, accordingly, the concern about endogeneity $E(X_{i}^{\mathrm{obs}}\varepsilon
_{i})\neq0$ would never arise. In other words, whenever instrumental
variable methods are invoked, it is fundamental to understand what feature
or interpretation of (\ref{textbook}) distinguishes it from the statistical
regression equation, and for what reason we should suspect the
dependence of
$X_{i}^{\mathrm{obs}}$ and $\varepsilon_{i}$.

Having a simple example would help us answer these questions. Consider a
classical problem of estimation of a production function. $ Q$ denotes the
quantity of a homogeneous good produced and $L$ is the measure of labor
input used (e.g., total hours worked by the employees). We do not consider
control covariates for now. Assume that the production technology of
firm $%
i$ is given by the following function,
\begin{eqnarray*}
Q_{i}(L)=\exp(\beta_{0}+\alpha_{i})L^{\beta_{1}},\quad
0<\beta _{1}<1,
\end{eqnarray*}
where $\beta_{0}$ is an unknown constant, $\alpha_{i}$ is a mean zero
unobserved productivity of firm $i$, and $\beta_{1}$ is the parameter of
interest assumed to be constant across firms. The specified production
function leads to a log-linear equation,
%
\begin{equation}\label{production}
Y_{i}(x)=\beta_{0}+\beta_{1}x+
\alpha_{i},
\end{equation}
where $x=\log L$ and $Y_{i}(x)=\log Q_{i}(L)$. This equation can be indeed
interpreted as the causal relationship between output and input in the
production process of firm $i$. As in equation (3.3) of the Imbens'
article, $Y_{i}(x)$ can be interpreted as $i$'s potential outcomes at each
possible input level $x\in\mathcal{X}$. In econometrics terminology,
equation (\ref{production}) is interpreted as a \textit{structural equation}
in the sense that it can generate any counterfactual outcomes of unit $i$
with respect to any manipulations in $x$. Note that the structural
equation (\ref{production}) relies only on the assumption or knowledge about
the underlying causal mechanism (production function) and, so far, no
considerations on how the data are generated have entered our
discussion yet.

Suppose that available data of pairs of log-output and log-input of $n$
producers, $ ( Y_{i}^{\mathrm{obs}},X_{i}^{\mathrm{obs}} ) $, $i=1,\dots,n$, are
observational, meaning that the observed input level $X_{i}^{\mathrm{obs}}$ can be
seen as a ``choice'' made by a firm~$i$. Following \citet{MarschakAndrews1944}, let us model each firm's choice of $X$ based on the following three
assumptions, (1) firms are \textit{rational}, meaning that each firm chooses
its input to maximize own profit, (2)~the market is under \textit{perfect
competition}, implying that every firm treat prices of the good and input
(wage) as given and (3) firms have complete knowledge of their production
technologies $\beta_{0}$, $\beta_{1}$ and $\alpha_{i}$ when they choose
their input levels. Under these somewhat unrealistic assumptions, firm
$i$%
's input choice solves the following profit maximization problem:
\begin{eqnarray*}
X_{i}^{\mathrm{obs}} =\log L_{i}^{\mathrm{obs}},
\end{eqnarray*}
 where
\begin{eqnarray*}
 L_{i}^{\mathrm{obs}} =\arg\max_{L}
\bigl\{ pQ_{i} ( L ) -w_{i}L_{i} \bigr\},
\end{eqnarray*}
where $p$ is the (common) price of the good, and $w_{i}$ is the hourly wage
given to firm $i$, which can vary over $i$, that is, the wage is
determined at
a localized labor market. The resulting choice $X_{i}^{\mathrm{obs}}$ is%
%
\begin{equation}\label{Xiobs}
X_{i}^{\mathrm{obs}}=\frac{1}{1-\beta_{1}} \biggl[ \beta_{0}+\log
\biggl( \frac
{p\beta
_{1}}{w_{i}} \biggr) +\alpha_{i} \biggr] .
\end{equation}
If we replace $x$ with $X_{i}^{\mathrm{obs}}$ in (\ref{production}) and notate $%
Y_{i}^{\mathrm{obs}}=Y_{i} ( X_{i}^{\mathrm{obs}} ) $, we obtain%
%
\begin{equation}\label{productionstructural}
Y_{i}^{\mathrm{obs}}=\beta_{0}+\beta_{1}X_{i}^{\mathrm{obs}}+
\alpha_{i} .
\end{equation}
This equation coincides with an equation of the form (\ref{textbook})
without covariates. Equation (\ref{Xiobs}) says that a more productive
(higher $\alpha_{i}$) firm chooses a larger labor input, implying that
the endogeneity problem $E ( X_{i}^{\mathrm{obs}}\alpha_{i} ) \neq0$ is
present. Accordingly, (\ref{productionstructural}) must differ from the
linear projection equation of $Y_{i}^{\mathrm{obs}}$ onto $X_{i}^{\mathrm{obs}}$, and the
least squares regression of $Y_{i}^{\mathrm{obs}}$ onto $X_{i}^{\mathrm{obs}}$ fails to
consistently estimate $\beta_{1}$. Here, the keypoints are (1) there
is a
specific causal model (\ref{production}) underlying (\ref{productionstructural}), and (2) the subject's optimal ``choice'' based on the
unobservable (to data analysts) causes correlation $E(X_{i}^{\mathrm{obs}}\alpha
_{i})\neq0$.

What can be a reasonable instrumental variable in the
current example? A
search for an instrumental variable can also be model-based. For instance,
if $w_{i}$ is available in data, equation (\ref{Xiobs}) says that $%
X_{i}^{\mathrm{obs}}$ should be dependent on $w_{i}$, while structural equation
(\ref%
{production}) says $w_{i}$ does not directly affect the output; accordingly,
$w_{i}$ satisfies the instrument relevance and the instrument exclusion
restriction. The validity of random assignment $E(w_{i}\alpha_{i})=0$, on
the other hand, would be questionable. For instance, firms located in an
urban area can be more productive (higher $\alpha_{i}$) than those located
in a rural area, and the wage level in urban area can be higher than the
wage level in rural, possibly due to a higher living cost and availability
of more skilled labor force. The motivation for using control
covariates $%
V_{i}$ (e.g., a demeaned indicator of whether firm $i$ is located in an urban
area or in a rural area) is to cope with potential confounders of $w_{i}$
and $\alpha_{i}$. Following the way in which Imbens treats covariates
(Section~6), we assume conditional random assignment $w_{i}\perp\alpha
_{i}|V_{i}$, and specify the dependence of $\alpha_{i}$ and $V_{i}$ as
%
\begin{equation}\label{controls}
\alpha_{i}=\beta_{2}V_{i}+\varepsilon_{i}
\quad \mbox{with }\varepsilon _{i}\perp V_{i}.
\end{equation}
Here, $\varepsilon_{i}$ is firm $i$'s unobserved productivity measured
relative to conditional mean $E(\alpha_{i}|V_{i})$. Note that coefficient
parameter $\beta_{2}$ summarizes the dependence of $\alpha_{i}$ and
$V_{i}$%
, and we are not attaching a causal interpretation to $\beta_{2}$.
Plugging $\alpha_{i}$ into (\ref{productionstructural}) yields the
textbook setup of the linear instrumental variable model (\ref{textbook}),
for which the two stage least squares procedure yields a consistent
estimator for $ ( \beta_{0},\beta_{1},\beta_{2} ) $. As is
clear through this simple example, the textbook equation (\ref{textbook})
can be seen as a \textit{composite} of the causal (structural) equation
(\ref%
{productionstructural}) and the statistical dependence relationship
(\ref%
{controls}).

\section{Point Estimate Versus Bounds: A~Treatment Choice
Perspective}

The discussed paper also reviews the current debate about the meaningfulness
of the complier's causal effect (Section~4.6). Imbens advocates the
importance and practical values of reporting the complier's causal effect
for the reason that it is the only causal estimand point-identified under
the maintained assumptions. Imbens, at the same time, acknowledges that
the population average causal effect is a parameter of primary interest in
many contexts of causal inference, and he recommends to report also the
bounds of the population average causal effect. In my opinion, Imbens'
proposal is quite sensible if the main task of the data analyst is to make
``scientific reporting'' about the causal effects. The point-identified
causal parameter for compliers and the set-identified causal parameter for
the entire population reflect (partially) distinct aspects of the data
distribution, and, importantly, the best we can learn from data under
the maintained assumptions are only those.

The objectives of causal
studies are not only for ``scientific reporting,''
but also for assisting ``decision making'' of a policy maker. If the latter
is a main task of the data analyst, then my personal view is that
neither of
the complier's causal effect estimate nor the bounds of the average causal
effect should be the final output that the decision maker would find most
useful. To make my argument more concrete, suppose that the decision
maker's objective is to maximize the social welfare defined by the sum of
individual outcomes over the target population. As in \citet{Chamberlain2011},
we suppose that he/she solves the treatment choice problem based on a
posterior belief for the social welfare, that is, the decision maker is
Bayesian. Since a comparison of the social welfare between the cases with
and without implementation of the treatment depends only on the population
average causal effect, the posterior distribution of the average causal
effect obtained from her/his carefully specified prior input leads to the
decision maker's optimal choice (see \cite{ChickeringandPearl1997} and \cite{ImbensRubin1997}) for Bayesian estimation of the average causal effect). On
the other hand, point estimates and inferential statements for the
complier's causal effect and the bounds for the population causal
effect do
not directly guide formal decision-making.

The argument I just gave crucially relies on the Bayesian premise that the
decision maker can fully specify a prior for the potential outcomes
distributions. This may not always be the case depending on a context.
Given the absence of a universal consensus on a ``noninformative'' prior,
inability to specify a credible prior becomes a serious concern especially
when the causal effect of interest is not identified, since the lack of
identification makes the posterior sensitive to a choice of prior no matter
how large the sample size is. One way to overcome this practical
difficulty would be to follow Manski's (\citeyear{Manski2000}, \citeyear{Manski2005}) frequentist approach
based on the minimax and minimax regret decision principle, which relies
only on the knowledge of the bounds of the population average causal
effect.

The Bayesian approach and Manski's data-alone approach are each
grounded in the two extreme schools of statistics. This means that there
should certainly be a room for blending the aspects of these two approaches
to complement their advantages and disadvantages. One compromising
approach would be to perform a minimax or minimax-regret decision analysis
with multiple priors/posteriors, namely, the $\Gamma$-minimax or
$\Gamma$%
-minimax regret decision analysis (see, e.g., \cite*{Berger1985}, Chapter~4).
For instance, in the current context, we can consider constructing a
\textit{%
set of posteriors} of average causal effects by combining a single
\textit{posterior} for the
identifiable parameters (causal effects for compliers, the mean of treatment
outcome for always-takers, the mean of control outcome for never-takers)
with a \textit{collection of priors} of the nonidentified parameters (the
mean of control outcome for always-takers and the mean of treatment outcome
for never-takers). The collection of priors for the nonidentified
parameters may represent the decision maker's partial or vague prior
knowledge, or represent the degree of robustness that the decision maker
wants to maintain in making the decision. Here, a single prior for the
identified parameters would make sense in a scenario that the decision maker
feels less anxious about a prior mis-specification for the identifiable
parameters since he/she knows data will well update it. If the class of
priors for the nonidentified parameters is not as large as the one that
allows for arbitrary ones, the resulting posterior $\Gamma$-minimax
treatment choice
rule will not be as conservative as the Manski's data-alone minimax
treatment choice rule based solely on the bounds. At the same time, unlike
the standard Bayesian analysis with a single prior distribution, it can
lead to a decision-making with taking
into account the posterior sensitivity concern with respect to a choice
of a prior for the
nonidentified parameters.
%

\section*{Acknowledgments}
I gratefully acknowledge the financial supports received from the ESRC
through the ESRC Centre for Microdata Methods and Practice (CeMMAP) (grant
number RES-589-28-0001).




\begin{thebibliography}{9}


\bibitem[\protect\citeauthoryear{Angrist, Imbens and Rubin}{1996}]{AIR}
\begin{barticle}[auto:STB|2014/06/18|12:29:53]
\bauthor{\bsnm{Angrist},~\bfnm{J.}\binits{J.}},
\bauthor{\bsnm{Imbens},~\bfnm{G.}\binits{G.}} \AND
\bauthor{\bsnm{Rubin},~\bfnm{D.}\binits{D.}}
(\byear{1996}).
\btitle{Identification of causal effects using instrumental variables}.
\bjournal{J. Amer. Statist. Assoc.}
\bvolume{91}
\bpages{444--455}.
\end{barticle}
\bptok{imsref}%
\endbibitem

\bibitem[\protect\citeauthoryear{Berger}{1985}]{Berger1985}
\begin{bbook}[mr]
\bauthor{\bsnm{Berger},~\bfnm{James~O.}\binits{J.~O.}}
(\byear{1985}).
\btitle{Statistical Decision Theory and {B}ayesian Analysis},
\bedition{2nd} ed.
\bpublisher{Springer},
\blocation{New York}.
\bid{doi={10.1007/978-1-4757-4286-2}, mr={0804611}}
\end{bbook}
\bptok{imsref}%
\endbibitem

\bibitem[\protect\citeauthoryear{Chamberlain}{2011}]{Chamberlain2011}
\begin{bincollection}[auto:STB|2014/06/18|12:29:53]
\bauthor{\bsnm{Chamberlain},~\bfnm{G.}\binits{G.}}
(\byear{2011}).
\btitle{Bayesian aspects of treatment choice}.
In \bbooktitle{The Oxford Handbook of Bayesian Econometrics}
(\beditor{\bfnm{J.}\binits{J.}~\bsnm{Geweke}},
\beditor{\bfnm{G.}\binits{G.}~\bsnm{Koop}} \AND
\beditor{\bfnm{H.}\binits{H.}~\bparticle{van}~\bsnm{Dijk}}, eds.).
\bpublisher{Oxford Univ. Press},
\blocation{Oxford}.
\end{bincollection}
\bptok{imsref}%
\endbibitem



\bibitem[\protect\citeauthoryear{Chickering and Pearl}{1997}]{ChickeringandPearl1997}
\begin{barticle}[mr]
\bauthor{\bsnm{Chickering},~\bfnm{D.}\binits{D.}} \AND
\bauthor{\bsnm{Pearl},~\bfnm{Judea}\binits{J.}}
(\byear{1997}).
\btitle{A clinician's tool for
analyzing non-compliance}.
\bjournal{Computing Science and Statistics}
\bvolume{29}
\bpages{424--431}.
\bid{doi={10.1007/978-1-4612-1842-5_3}, mr={1601275}}
\end{barticle}
\bptok{imsref}%
\endbibitem



\bibitem[\protect\citeauthoryear{Imbens and Angrist}{1994}]{ImbensandAngrist1994}
\begin{barticle}[auto:STB|2014/06/18|12:29:53]
\bauthor{\bsnm{Imbens},~\bfnm{G.}\binits{G.}} \AND
\bauthor{\bsnm{Angrist},~\bfnm{J.}\binits{J.}}
(\byear{1994}).
\btitle{Identification and estimation of local average treatment effects}.
\bjournal{Econometrica}
\bvolume{62}
\bpages{467--475}.
\end{barticle}
\bptok{imsref}%
\endbibitem

\bibitem[\protect\citeauthoryear{Imbens and Rubin}{1997}]{ImbensRubin1997}
\begin{barticle}[mr]
\bauthor{\bsnm{Imbens},~\bfnm{Guido~W.}\binits{G.~W.}} \AND
\bauthor{\bsnm{Rubin},~\bfnm{Donald~B.}\binits{D.~B.}}
(\byear{1997}).
\btitle{Bayesian inference for causal effects in
randomized experiments with noncompliance}.
\bjournal{Ann. Statist.}
\bvolume{25}
\bpages{305--327}.
\bid{mr={1429927}}
\end{barticle}
\bptok{imsref}%
\endbibitem

\bibitem[\protect\citeauthoryear{Manski}{2000}]{Manski2000}
\begin{barticle}[auto:STB|2014/06/18|12:29:53]
\bauthor{\bsnm{Manski},~\bfnm{C.}\binits{C.}}
(\byear{2000}).
\btitle{Identification problems and decisions under ambiguity: Empirical analysis of treatment response and normative analysis of treatment choice}.
\bjournal{J. Econometrics}
\bvolume{95}
\bpages{415--442}.
\end{barticle}
\bptok{imsref}%
\endbibitem

\bibitem[\protect\citeauthoryear{Manski}{2005}]{Manski2005}
\begin{bbook}[mr]
\bauthor{\bsnm{Manski},~\bfnm{Charles~F.}\binits{C.~F.}}
(\byear{2005}).
\btitle{Social Choice with Partial Knowledge of Treatment Response}.
\bpublisher{Princeton Univ. Press},
\blocation{Princeton, NJ}.
\bid{mr={2178946}}
\end{bbook}
\bptok{imsref}%
\endbibitem

\bibitem[\protect\citeauthoryear{Marschak and Andrews}{1944}]{MarschakAndrews1944}
\begin{barticle}[mr]
\bauthor{\bsnm{Marschak},~\bfnm{Jacob}\binits{J.}} \AND
\bauthor{\bsnm{Andrews},~\bfnm{William~H.}\binits{W.~H.} \bsuffix{Jr.}}
(\byear{1944}).
\btitle{Random simultaneous equations and the theory of production}.
\bjournal{Econometrica}
\bvolume{12}
\bpages{143--205}.
\bid{issn={0012-9682}, mr={0011941}}
\end{barticle}
\bptok{imsref}%
\endbibitem



\end{thebibliography}
\end{document}